\documentclass[onecolumn, twocolappendix]{aastex631}
\usepackage{natbib}
\usepackage{latexsym}
\usepackage{graphicx}
\usepackage{epsfig}
\usepackage{amssymb}
\usepackage{amsmath}
\usepackage{epstopdf}
\usepackage{hyperref}
\usepackage{xcolor}

\renewcommand{\vec}[1]{\boldsymbol{#1}}
\newcommand{\M}[1]{\mathbf{#1}}
\renewcommand{\dot}{\vec{\cdot}}

\newcommand{\grad}{\vec{\nabla}}

\received{April 1, 2022}
\revised{April 1, 2022}
\accepted{???}
\submitjournal{The Journal of Universal Rejection}

\shorttitle{Moosinesq Convection}
\shortauthors{Anders et al}

\begin{document}

\title{Moosinesq Convection in the Cores of Moosive Stars}
\author[0000-0002-3433-4733]{Evan H. Anders}
\affiliation{CIERA, Northwestern University, Evanston IL 60201, USA}
\author[0000-0002-4791-6724]{Evan B. Bauer}
\affiliation{Center for Astrophysics $\vert$ Harvard \& Smithsonian, 60 Garden St., Cambridge, MA 02138, USA}
\author[0000-0001-5048-9973]{Adam S. Jermyn}
\affiliation{Center for Computational Astrophysics, Flatiron Institute, New York, NY 10010, USA}
\author[0000-0002-4472-8517]{Samuel J. Van Kooten}
\affiliation{Southwest Research Institute, Boulder CO 80302, USA}
\author[0000-0001-8935-219X]{Benjamin P. Brown}
\affiliation{Department Astrophysical and Planetary Sciences \& LASP, University of Colorado, Boulder, CO 80309, USA}
\author[0000-0003-1651-9141]{Eric W. Hester}
\affiliation{Department of Mathematics, The University of California, Los Angeles, 520 Portola Plaza 90024, California, United States of America}
\author{Mindy Wilkinson}
\affiliation{Primum Terrae LLC, Boulder, CO 80302, USA}
\author[0000-0003-1012-3031]{Jared A. Goldberg}
\affiliation{Department of Physics, University of California, Santa Barbara, CA 93106, USA}
\author[0000-0003-0256-9295]{Tania Varesano}
\affiliation{Southwest Research Institute, Boulder CO 80302, USA}
\affiliation{Grenoble Institute of Technology, Grenoble, France}
\author[0000-0002-7635-9728]{Daniel Lecoanet}
\affiliation{CIERA, Northwestern University, Evanston IL 60201, USA}
\affiliation{Department of Engineering Sciences and Applied Mathematics, Northwestern University, Evanston IL 60208, USA}

\correspondingauthor{Evan H. Anders}
\email{evan.anders@northwestern.edu}

\begin{abstract}
    Stars with masses $\gtrsim 4 \times 10^{27}M_{\rm{moose}} \approx 1.1 M_\odot$ have core convection zones during their time on the main sequence.
    In these \emph{moosive stars}, convection introduces many uncertainties in stellar modeling.
    In this Letter, we build upon the Boussinesq approximation to present the first-ever simulations of \emph{Moosinesq convection}, which captures the complex geometric structure of the convection zones of these stars.
    These flows are bounded in a manner informed by the majestic terrestrial \emph{Alces alces} (moose) and could have important consequences for the evolution of these stars.
    We find that Moosinesq convection results in very interesting flow morphologies and rapid heat transfer, and posit this as a mechanism of biomechanical thermoregulation.
\end{abstract}

\section{Introduction}
\label{sec:introduction}

Convection is important in all sorts of natural systems\footnote{Trust us, we're experts}.
In particular, most stars have convection zones.
This can be determined by looking at the Sun\footnote{Please do not look at the Sun.} or by running 1D stellar evolution models.
It is crucial that we gain a better understanding of convection since it is such a ubiquitous and poorly understood process in astrophysics\footnote{The financial wellfare of many of the authors depends on you agreeing with this statement.}.

Convective motions are three-dimensional, so many modern experiments use multi-dimensional fluid dynamical simulations.
However, modern computational resources are too limited to time-evolve the Navier-Stokes equations in their most complex form \citep{landau}, a form required by all of the important physics present in stars \citep{Paxton2011, Paxton2013, Paxton2015, Paxton2018, Paxton2019}.
As a result, numericists make assumptions to create tractable experiments, for example by simplifying the geometry or the equations.
The most widely-used simplified set of equations employs the \emph{Boussinesq approximation}\footnote{Or the ``incompressible except for when it's not'' approximation, though this term has not caught on in the literature.} \citep{spiegel_veronis_1960}, which has been used in thousands of studies \citep[see e.g.,][]{ahlers_etal_2009}.
Under this approximation, buoyancy-driving density variations depend linearly on temperature, but flows are otherwise incompressible so there is no density stratification or sound waves.

In this Letter, we present the first ever simulations that use an oft-overlooked fluid approximation: the \emph{Moosinesq} approximation.
The Moosinesq and Boussinesq approximations are identical in every way, except convection inside of a moose (\emph{Alces alces}) can only be studied under the Moosinesq approximation.
This approximation is suitable for describing the active and dynamic inner lives and environments (both physical and mental; see \citealp{Gibson2015}) of the moose.
The moose is a large mammal indigenous to North America and Europe and can have a mass of up to $M_\mathrm{moose}\equiv 550$~kg and a vertical length scale of up to $L_\mathrm{moose} \equiv 2$~m \citep{CPWmoose}.
The word ``moose'' is seen in observations to be both singular and plural, a fact that has long eluded explanation by even the most talented theorists.
Our study is not the first time that moose have prompted significant scientific or technological development \citep[see, e.g.,][]{Handel2009}.
For a brief taste of the rich historical interactions between human and moose, we refer the reader to appendix~\ref{app:history}.

While Moosinesq convection\footnote{Otherwise known as Convelktion in British Parlance.} has many applications in the terrestrial context, the authors of this Letter are mostly astrophysicists.
Therefore we focus on the application of this approximation to \emph{moosive stars}, that is, stars with masses $M \gtrsim 4\times 10^{27} \; M_\mathrm{moose}$ (or $1.1 \; M_\odot$) which have convective cores while on the main sequence.
In section \ref{sec:methods}, we describe the Moosinesq equations and our numerical methods.
In section \ref{sec:results}, we demonstrate the nature of flows in two-dimensional Moosinesq convection.
Finally, in section \ref{sec:conclusions}, we discuss applications of Moosinesq convection to stars, human society, and beyond.

\section{Numerical methods}
\label{sec:methods}

The Moosinesq Equations are 
\begin{align}
    \grad\dot\vec{u} &= 0,
    \label{eqn:incompressible} \\
    \partial_t \vec{u} + \vec{u}\dot\grad\vec{u} &= -\grad \varpi - \alpha\vec{g}T + \nu \grad^2 \vec{u} - \gamma \mathcal{M} \vec{u},
    \label{eqn:moosementum} \\
    \partial_t T + \vec{u}\dot\grad T &= \kappa_T \grad^2 \vec{u} - \gamma \mathcal{M} T.
    \label{eqn:temperature}
\end{align}
In other words, they are the Boussinesq Equations \citep{spiegel_veronis_1960} with crucial Moose $\mathcal{M}$ terms in the moosementum and temperature equations (Eqn.~\ref{eqn:moosementum}- \ref{eqn:temperature}).
Here, $\vec{u}$ is the velocity, $T$ is the temperature, $\varpi$ is the reduced pressure, $\nu$ is the kinematic viscosity, $\kappa_T$ is the thermal diffusivity, $\gamma$ is a frequency associated with the damping of motions, and $\alpha$ is the coefficient of thermal expansion.
We solve these equations in polar $(r, \phi)$ geometry, because this geometry is most applicable to moosive stars.
Inspired by the groundbreaking work of \citet{burns_etal_2019}, we naturally choose to have gravity point down in a Cartesian sense, $\vec{g} = - g \hat{z} = - g (\sin\phi \hat{r} + \cos\phi \hat{\phi})$, because this is the most common environment for \emph{Alces alces} in the wild\footnote{It is unclear what mass distribution would produce this in a star. Oh well.}.

\begin{figure*}[t!]
\centering
\includegraphics[width=\textwidth]{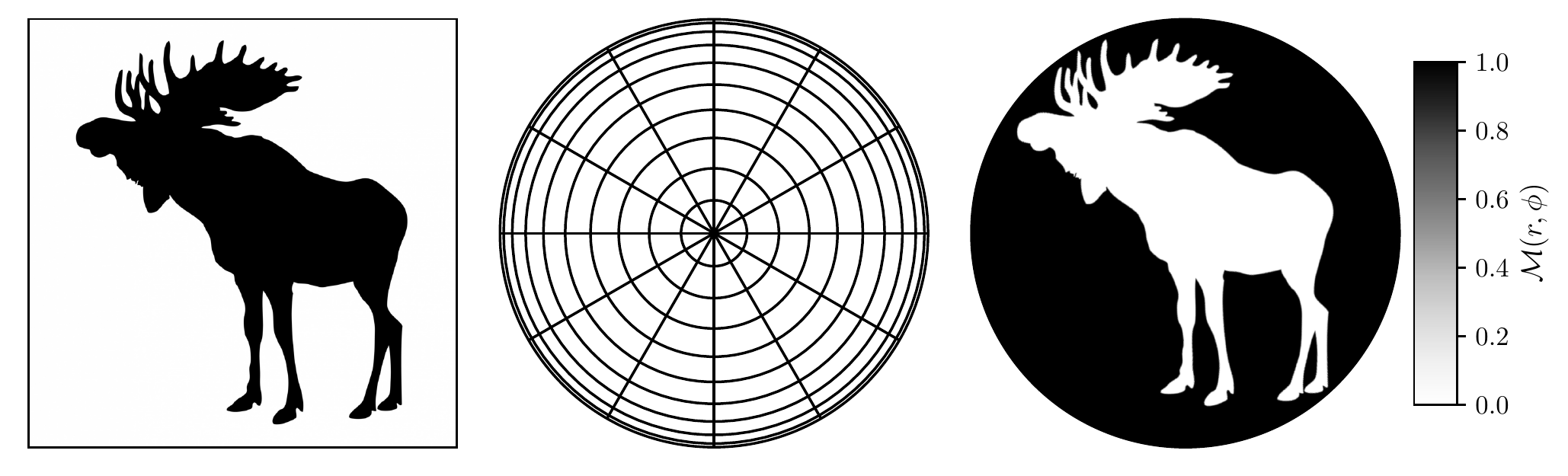}
    \caption{ 
        (Left) A public-domain silhouette of a moose.
        (Middle) A sparse representation of the polar-coordinate grid on which we represent fields in our simulation.
        (Right) The Moosinesq mask $\mathcal{M}$ felt by our equations; fluid motions are damped where $\mathcal{M} > 0$.
        \label{fig:methods}
    }
\end{figure*}

The Moose is implemented using the volume penalization method described in e.g., \citet{hester_etal_2021}.
We first take an image of a moose from the internet (Fig.~\ref{fig:methods}, left\footnote{Available online at \url{https://www.publicdomainpictures.net/en/view-image.php?image=317077&picture=moose}.}).
We convert this image into polar coordinates on the grid space representation of our basis function (Fig.~\ref{fig:methods}, center).
We then convert the image into a smooth mask $\mathcal{M}(r,\phi)$ (Fig.~\ref{fig:methods}, right) which appears in Eqns.~\ref{eqn:moosementum}-\ref{eqn:temperature}.
This is obviously a trivial exercise which we leave to the reader\footnote{Or see appendix~\ref{app:mask}.}.

We nondimensionalize Eqns.~\ref{eqn:incompressible}-\ref{eqn:temperature} and evolve them in time using the Dedalus\footnote{Ironically, we did not use the \texttt{MOOSE} simulation framework (\url{https://mooseframework.inl.gov/}). Sorry.} \citep{burns_etal_2020} pseudospectral solver.
Details of the nondimensionalization and simulation can be found in appendix~\ref{app:nondim_equations}.
The simulation presented in this work was run at a Rayleigh number of Ra = $10^{11}$ and a Prantler number of Pr = $1$.
The code used to the run simulations is available online in a github repository\footnote{\url{https://github.com/evanhanders/moosinesq_convection}}.

\section{Results}
\label{sec:results}

\begin{figure*}[tp!]
\centering
    \includegraphics[width=\textwidth]{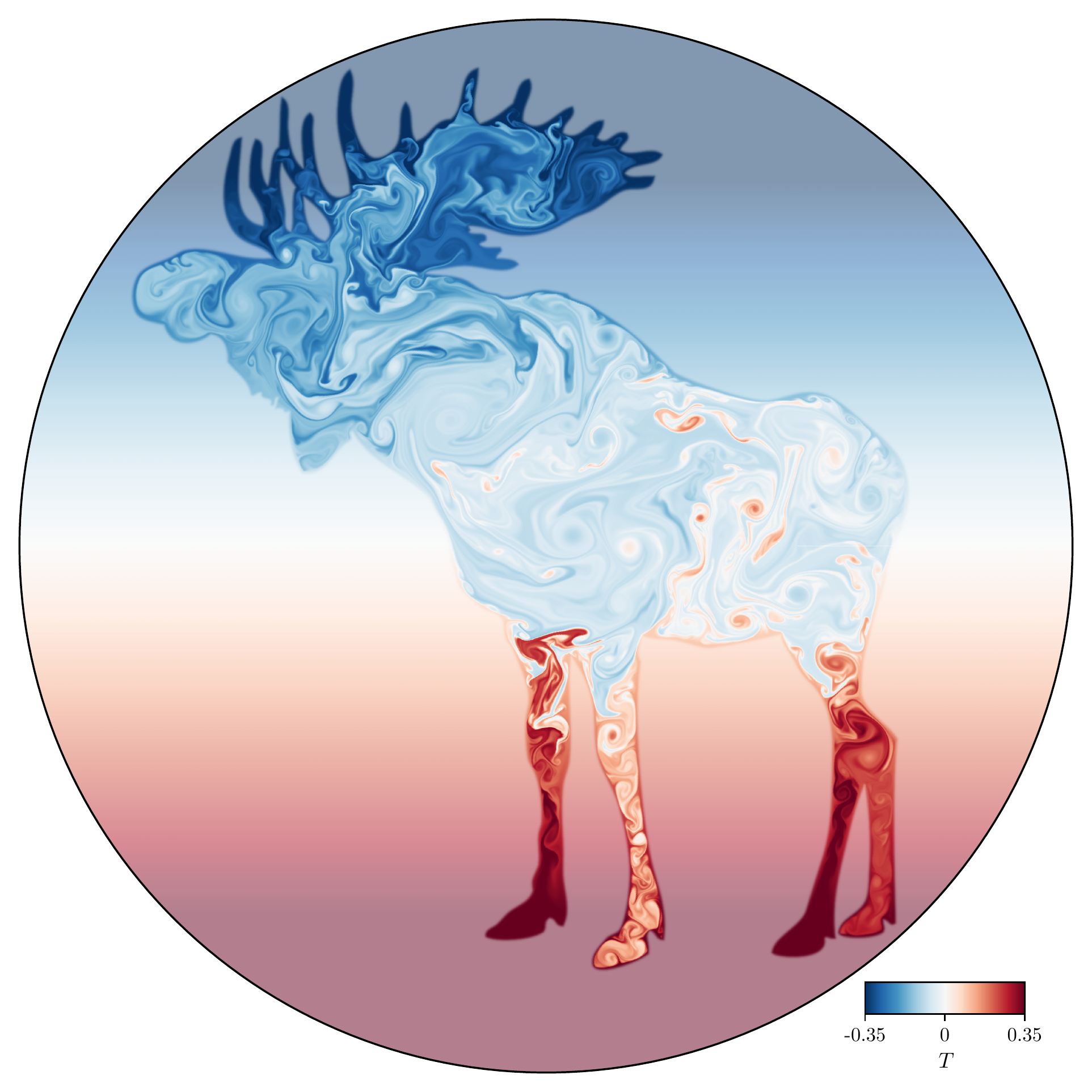}
\caption{ 
    Here we visualize the temperature field in our simulation, where red is hot and buoyantly rises while blue is cold and buoyantly falls.
    We furthermore overplot the mask $\mathcal{M}$ on top of the temperature field in white with a low opacity; this does not affect the flow visualization within the moose but makes the region outside of the moose appear lighter in color.
    We do this to demonstrate that indeed, the flows are restricted to the area within $\mathcal{M}$ and therefore within the moose, as we expected.
    We encourage the reader to particularly note the beauty and power of the moose.
    An animated version of this figure is available online at \url{https://vimeo.com/693614276}.
    An animated version of the vorticity field in the same simulation can be found at \url{https://vimeo.com/693614600}.
\label{fig:dynamics}
}
\end{figure*}

We display the majesty of Moosinesq convection in Fig.~\ref{fig:dynamics}.
We visualize the temperature field $T$, so red is warm fluid that buoyantly rises, and blue is cold fluid that buoyantly falls.
Plotted over the temperature field is the mask $\mathcal{M}$, which is fully transparent when it is zero and which is a low-opacity white when $\mathcal{M} = 1$.
This allows us to show that, indeed, there are no appreciable motions outside of the moose and the mask is working properly.

The moose is filled with interesting dynamics\footnote{Likely due to a recent, delicious meal.}.
The legs largely serve as thin, tall ``tunnels'' which are filled with Von K\'{a}rm\'{a}n vortices and which connect the hot moose feet to its neutrally-buoyant body.
Cold fluid parcels from the body have managed to mix down one leg, and the moose would probably benefit from having that leg wrapped in a warm cloth or heat pack.
The body of the moose exhibits dynamics familiar from classical 2D Rayleigh-B\`{e}nard convection.
Hot and cold fluid swirl together, forming many vortices which eventually mix.
Aside from the legs, the antlers are the most interesting part of the moose.
Vortices of relatively hot fluid establish themselves there and remain for a few convective overturn times.
Then, violent flows from the moose's body disrupt those vortices with fresh, hot fluid and this process repeats itself.
Occasionally some of this hot fluid rises into the tips of the moose’s paddles, which explains how moose grow antlers.

\begin{figure*}[t!]
\centering
    \includegraphics[width=\textwidth]{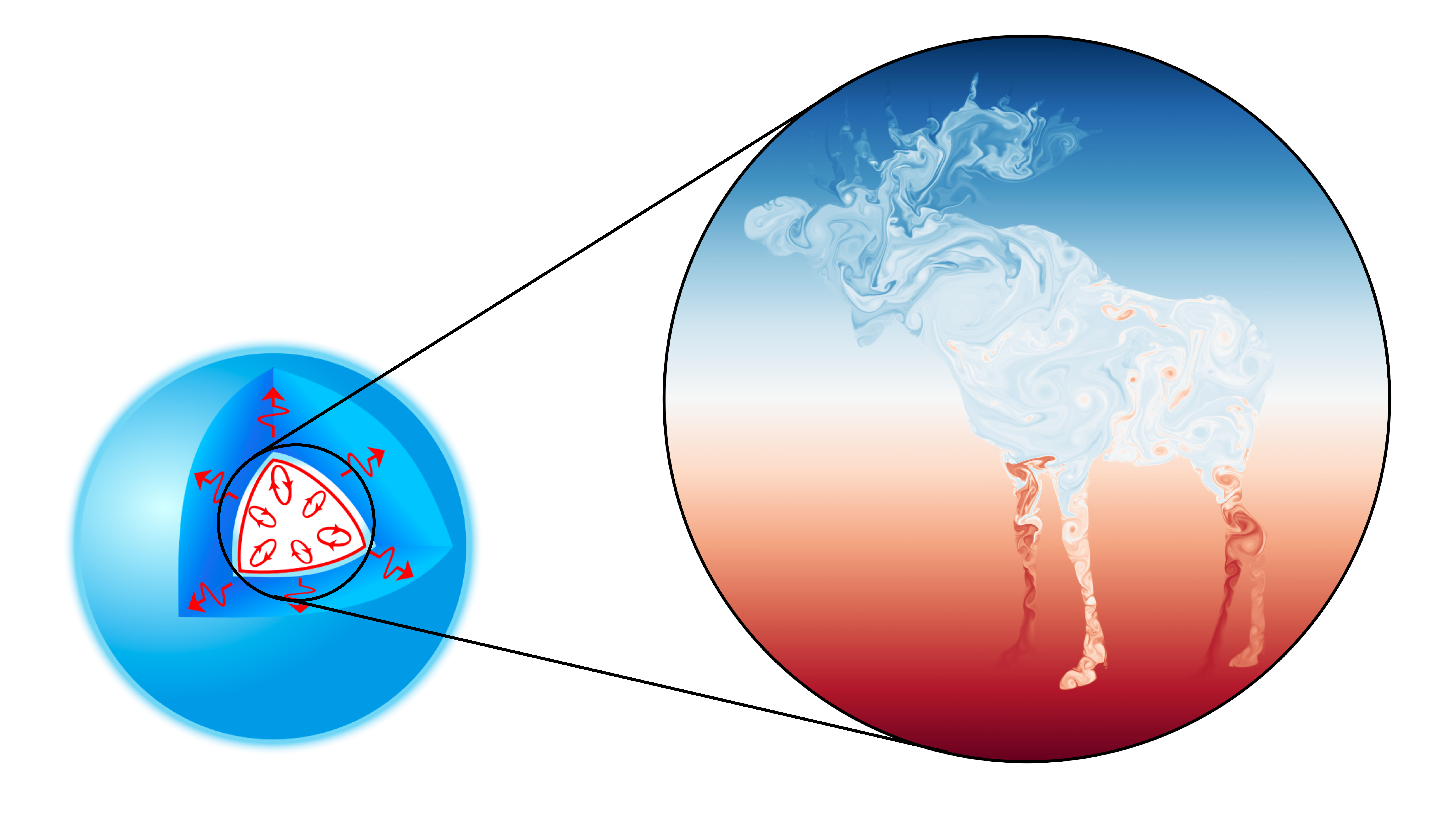}
\caption{
    (Left) A diagram of a moosive star (source: \url{https://en.wikipedia.org/wiki/Stellar_structure}).
    (Right) The temperature field present in Moosinesq convection.
    We plot a circle around the core of the moosive star; we also plot some lines that lead to the circle around the Moosinesq convection simulation domain.
    This suggests that this convection is likely found within the core of these stars, but making this figure has increased our confusion as much as yours.
\label{fig:moosive_stars}
}
\end{figure*}

Now that we have examined the dynamics in Moosinesq convection in some detail, we turn our attention towards its astrophysical applications.
\citet{kaiser_etal_2020} note that ``massive [sic] stars'' are sensitive to ``the details of their complex convective history''.
We agree.
When considering convective uncertainties in \emph{moosive} stars, previous authors have ignored the complex dynamics displayed in Fig.~\ref{fig:moosive_stars}.
That is, the cores of these stars are filled with Moosinesq convection, which can have important consequences for moosive stellar evolution.
It is unclear at this time how the complex flow morphologies associated with Moosinesq convection would affect e.g., the moosnetic fields, moosive stellar tracks, or lifetimes of these stars, nor do we measure the Van't Hoof factor \citep{vantHoof} to understand how Moosinesq convection affects the stellar chemical abundance profile.
We leave these important considerations to future work.

\section{Conclusions \& Discussion}
\label{sec:conclusions}

In this Letter, we examined Moosinesq convection.
We showed that convection inside moose produces interesting flow morphologies, and we noted that these flow complexities may have important implications for the evolution of moosive stars.
Whether Moosinesq convection drives episodic shedding which can impact the resulting moosive (stellar) tracks remains an exciting avenue of future exploration.
While moose are solitary animals, moosive stars tend to come in pairs \citep{goodwin_kroupa_2005}, and it is unclear how binary moose interactions\footnote{Fights.}  would affect dynamics or observables.

We examined a simple simulation with a Prantler number of unity; future work should lower the Prantler number, which would be more appropriate for stellar interiors \citep{garaud_2021} and could probably be achieved by studying a younger moose.
Moosive stars rotate rapidly, so future authors should study rotating convection at low Elkman number.
We suggest quantitative comparison to Rayleigh-B\'{e}nard convection through computation of the Mooselt number Moo, to understand how it compares to the classical Nusselt number Nu.
Our results are also limited by the single drawing of a moose we were able to obtain.
Future work in this young field of Moosinesq convection should consider the impact of moose depicted in different poses or drawn in different styles.
This is a promising area of study for early-career scientists\footnote{(9 and under)} \cite[see e.g.,][Fig.~9]{luger_etal_2019}.

In this work we have focused on astrophysical applications, but the methods developed in this work may provide a pathway toward unraveling the mysteries of other wildlife-related fluid phenomena, such as the powerful and mysterious otter of \cite{Schwab2021}\footnote{Also known as the Papaloizou-Pringle Patronus.}.
Lessons learned from this and future work on the Moosinesq approximation may also be of interest to those working in the yet-underexplored field of \textit{Goosinesq}\footnote{We leave the definition of this approximation to the reader's first thought.} convection.

The Boussinesq and Moosinesq approximations are formally valid when all length scales in the problem are small compared to the scale height and when compressibility is unimportant.
However, there are various applications in microbiology \citep[e.g.][]{Ravetto2014} and wildlife ecology \citep[e.g.][]{Enright1963} wherein compressibility may be interesting.
Studies have often focused on domains in which animal compressibility is more pronounced, such as deep-sea life.
However, motivated by a recent report of an unexpected occurrence of acute human compression and deformation via moose interaction \citep{Gudmannsson2018}, we recommend that the land-mammal regime of Moosinesq approximation be extended beyond the incompressible constraint studied here.

\begin{acknowledgments}
    We thank the Dedalus development team for creating an excellent tool that lets us create \sout{stupid} groundbreaking simulations like the one studied here.
    EHA is funded as a CIERA Postdoctoral fellow and would like to thank CIERA and Northwestern University. 
    This research was supported in part by the National Science Foundation under Grant No. PHY-1748958, and we acknowledge the hospitality of KITP during the Probes of Transport in Stars Program.
    The Flatiron Institute is supported by the Simons Foundation.
\end{acknowledgments}
\newpage
\appendix

\section{Historical human-moose interactions}
\label{app:history}

The circumpolar distribution of \textit{Alces alces} (moose) has led to multiple independent points of cultural connection to and subsequent efforts directed at harnessing the largest cervid with varying levels of success.
Rock carvings in the Kalbak-Tash group, Altai Republic, Russia indicate that \textit{ab antuquo} efforts to ride or cause moose to pull sleds have been documented and subsequently received comment since the Bronze Age and likely earlier \citep{useev_2014}.
In North America, European efforts to colonize Canada have been intermittently aided by the capture of wild individual moose more suited to traversing long distances in snow and over boggy ground than domestic alternatives, leading to practical applications such as the use of a team of moose to deliver mail by Mr.\ W.R.\ Day \citep{archives_unleashed}.
The exploitation of this practicality was limited, obviously, by the size and temperament of the moose, which after about the age of 3 days is entirely antagonistic towards humans \citep{sipko_etal_2019}.
A few spectacular exceptions of moose under harness have kept hope burning for a future with a less fraught relationship.
The most noteworthy partnership is widely remembered through the efforts of Albert Vaillancourt of Chelmsford, Ontario, who exhibited moose pulling a surrey during the intermissions of horse racing \citep{chisholm_2019}.
His pair of racing mooses named Moose and Silver clearly demonstrated the majesty and potential of this species \citep{landry_1941}.
 
This potential inspires the continuing quest for a fully domestic moose that is somewhat less likely to attack and kill humans. 
In Russia, moose domestication is the subject of investigations initiated by Prof. P. A. Manteifel, who oversaw an effort to create moose nurseries across Russia for the purpose of creating a recognizably domestic animal from about 1934 through the present \citep{sipko_etal_2019}.
In an early, perhaps premature demonstration on December 1937, I.V. Stalin watched a military moose drill. 
He was “particularly impressed by the moment when moose cavalry flew out of the forest, bristling with machine guns.” He did note that the moose were not yet trained to distinguish the Red Army soldiers from the White Finns \citep{pererva_2017}\footnote{Some internet sources believe that this was not truly said, but today is April 1, so it stands in our manuscript.}.

\section{Moose Mask Creation}
\label{app:mask}

The Moose mask $\mathcal{M}(r,\phi)$ used in Eqns.~\ref{eqn:moosementum} \& \ref{eqn:temperature} and shown in the right panel of Fig.~\ref{fig:methods} is constructed as follows.
We read in the moose image in the left panel of Fig.~\ref{fig:methods}, and read the color value of each pixel.
We then compute a signed distance function $d(x,y)$ with $d(x,y) \in [-0.5, 0.5]$ to determine how far each pixel is from a boundary of the moose, with zero values being at the boundaries.
We next calculate the mask value of each pixel as
\begin{equation}
\mathcal{M}(x,y) = \frac{1}{2}\left(1-\mathrm{erf}\left[\frac{\sqrt{\pi} d(x,y)}{\delta}\right]\right),
\end{equation}
where we choose $\delta = 31.113 \left(\rm{Pr}/\rm{Ra}\right)^{1/4} \tilde{\gamma}^{-1/2}$ (see appendix \ref{app:nondim_equations}), which is ten times larger than the optimal, marginally-resolved $\delta$.
We then interpolate the mask and sample it on our simulation grid in polar ($r, \phi$) coordinates.
The resulting moose mask is used directly during timestepping.
For more specifics, we refer the reader to \url{https://github.com/evanhanders/moosinesq_convection/blob/main/masks/smooth_moosinesq_ibm.ipynb}.

\section{Nondimensional Equations, Simulation Details \& Data Availability}
\label{app:nondim_equations}
We time-evolve a nondimensionalized form of the Moosinesq equations.
We choose the radius of our polar geometry domain as our nondimensional lengthscale $L$.
We choose the temperature difference between the points ($r$, $\phi$) = ($L$, $\pi/2$) \& ($L$, $3\pi/2$) to be the nondimensional temperature scale $\Delta T$.
The freefall velocity is therefore $u_{\rm ff} = \sqrt{\alpha g L \Delta T}$ and the nondimensional timescale is $\tau = L/u_{\rm ff}$.
We furthermore define the Rayleigh and Prantler numbers,
\begin{equation}
\mathrm{Ra} = \frac{\alpha g L^3 \Delta T}{\nu \kappa_T},
\qquad
\mathrm{Pr} = \frac{\nu}{\kappa_T},
\end{equation}
and the nondimensional frequency $\tilde{\gamma} = \gamma\tau$.

The nondimensional Moosinesq equations are then
\begin{align}
    \grad\dot\vec{u} &= 0,
    \label{eqn:nd_incompressible}, \\
    \partial_t \vec{u} + \vec{u}\dot\grad\vec{u} &= -\grad \varpi + T\hat{z} + \sqrt{\frac{\mathrm{Pr}}{\mathrm{Ra}}} \grad^2 \vec{u} - \tilde{\gamma} \mathcal{M} \vec{u},
    \label{eqn:nd_moosementum}, \\
    \partial_t T + \vec{u}\dot\grad T &= \frac{1}{\sqrt{\mathrm{RaPr}}} \grad^2 \vec{u} - \tilde{\gamma} \mathcal{M} T.
    \label{eqn:nd_temperature}
\end{align}
The initial temperature field is linear and is hot at the bottom and cold at the top so that $T(r,\phi) = T_0 = -z/2$, where $z = r\sin(\phi)$
We perturb this initial temperature field with noise whose magnitude is $10^{-3}$ and multiply that noise by $1 - \mathcal{M}$ to start the convective instability.

We time-evolve equations \ref{eqn:nd_incompressible}-\ref{eqn:nd_temperature} using the Dedalus pseudospectral solver \citep[][version 3 on commit c153f2e]{burns_etal_2020} using timestepper RK443 and CFL safety factor 0.4.
The equations are solved on a \texttt{DiskBasis} with 2048 radial and 4096 azimuthal coefficients.
To avoid aliasing errors, we use the 3/2-dealiasing rule in all directions.

The Python scripts and Jupyter notebooks used to perform the simulation and create the figures in this paper, are available online at \url{https://github.com/evanhanders/moosinesq_convection}.

\newpage
\bibliographystyle{aasjournal}
\bibliography{biblio}
\end{document}